\documentclass[aps,prb,reprint,preprintnumbers,superscriptaddress]{revtex4-1}
\usepackage[T1]{fontenc}
\usepackage{times}
\usepackage{lettrine}
\usepackage{graphicx,color}
\usepackage{amsfonts,amsmath,amssymb,amsbsy}
\usepackage{lineno}
\usepackage{fancyhdr}
\usepackage{float}
\definecolor{orange}{rgb}{1,0.41,0.13}
\usepackage[colorlinks=true, linkcolor=blue, citecolor=blue, urlcolor=blue]{hyperref}
\def\section#1{\medskip\noindent\textbf{\large{#1}}\par}
\makeatletter
\renewcommand{\fnum@figure}{\figurename~\textbf{\thefigure}}
\makeatother
\renewcommand{\figurename}{\textbf{Fig.}}

\let\mathbf=\boldsymbol
\let\Gamma=\varGamma
\let\Upsilon=\varUpsilon
\def\emph#1{\textcolor{red}{#1}}

\begin{document}

\title{Bimeron Clusters in Chiral Antiferromagnets}

\author{Xiaoguang Li}
\thanks{These authors contributed equally to this work.}
\affiliation{School of Science and Engineering, The Chinese University of Hong Kong, Shenzhen, Guangdong 518172, China}
\affiliation{Hefei National Laboratory for Physical Sciences at the Microscale, Department of Physics, University of Science and Technology of China, Hefei 230026, China}

\author{Laichuan Shen}
\thanks{These authors contributed equally to this work.}
\affiliation{School of Science and Engineering, The Chinese University of Hong Kong, Shenzhen, Guangdong 518172, China}

\author{Yuhao Bai}
\thanks{These authors contributed equally to this work.}
\affiliation{Research Institute of Materials Science of Shanxi Normal University \& Collaborative Innovation Center for Shanxi Advanced Permanent Magnetic Materials and Technology, Linfen, Shanxi 041004, China}
\affiliation{School of Physics and Electronic Information, Shanxi Normal University, Linfen, Shanxi 041004, China}

\author{Junlin Wang}
\affiliation{Department of Physics, University of York, York YO10 5DD, United Kingdom}
\affiliation{York-Nanjing International Center of Spintronics (YNICS), Nanjing University, Nanjing 210093, China}

\author{Xichao Zhang}
\affiliation{School of Science and Engineering, The Chinese University of Hong Kong, Shenzhen, Guangdong 518172, China}
\affiliation{York-Nanjing International Center of Spintronics (YNICS), Nanjing University, Nanjing 210093, China}

\author{Jing Xia}
\affiliation{School of Science and Engineering, The Chinese University of Hong Kong, Shenzhen, Guangdong 518172, China}

\author{Motohiko Ezawa}
\affiliation{Department of Applied Physics, The University of Tokyo, 7-3-1 Hongo, Tokyo 113-8656, Japan}

\author{Oleg A. Tretiakov}
\affiliation{School of Physics, The University of New South Wales, Sydney 2052, Australia}

\author{Xiaohong Xu}
\affiliation{Research Institute of Materials Science of Shanxi Normal University \& Collaborative Innovation Center for Shanxi Advanced Permanent Magnetic Materials and Technology, Linfen, Shanxi 041004, China}
\affiliation{School of Chemistry and Materials Science of Shanxi Normal University \& Key Laboratory of Magnetic Molecules and Magnetic Information Materials of Ministry of Education, Linfen, Shanxi 041004, China}

\author{Michal Mruczkiewicz}
\affiliation{Institute of Electrical Engineering, Slovak Academy of Sciences, D\'ubravsk\'a cesta 9, 841 04 Bratislava, Slovakia}
\affiliation{Centre for Advanced Materials Application CEMEA, Slovak Academy of Sciences, D\'ubravsk\'a cesta 5807/9, 845 11 Bratislava, Slovakia}

\author{Maciej Krawczyk}
\affiliation{Faculty of Physics, Adam Mickiewicz University in Poznan, Uniwersytetu Poznanskiego 2, 61-614 Poznan, Poland}

\author{Yongbing Xu}
\affiliation{Department of Physics, University of York, York YO10 5DD, United Kingdom}
\affiliation{York-Nanjing International Center of Spintronics (YNICS), Nanjing University, Nanjing 210093, China}

\author{Richard F. L. Evans}
\affiliation{Department of Physics, University of York, York YO10 5DD, United Kingdom}

\author{Roy W. Chantrell}
\affiliation{Department of Physics, University of York, York YO10 5DD, United Kingdom}

\author{Yan Zhou}
\email[Email:~]{zhouyan@cuhk.edu.cn}
\affiliation{School of Science and Engineering, The Chinese University of Hong Kong, Shenzhen, Guangdong 518172, China}

\begin{abstract}\noindent
A mgnetic bimeron is an in-plane topological counterpart of a magnetic skyrmion. Despite the topological equivalence, their statics and dynamics could be distinct, making them attractive from the perspectives of both physics and spintronic applications. In this work, we investigate an antiferromagnetic (AFM) thin film with interfacial Dzyaloshinskii-Moriya interaction (DMI), and introduce the AFM bimeron cluster as a new form of topological quasi-particle. Bimerons demonstrate high current-driven mobility as generic AFM solitons, while featuring anisotropic and relativistic dynamics excited by currents with in-plane and out-of-plane polarizations, respectively. Moreover, these spin textures can absorb other bimeron solitons or clusters along the translational direction to acquire a wide range of N\'eel topological numbers. The clustering involves the rearrangement of topological structures, and gives rise to remarkable changes in static and dynamical properties. The merits of AFM bimeron clusters reveal a potential path to unify multi-bit data creation, transmission, storage and even topology-based computation within the same material system, and may stimulate innovative spintronic devices enabling new paradigms of data manipulations.
\end{abstract}

\date{\today}
\preprint{}

\maketitle

\section{Introduction}
\label{se:Introduction}

The last decade witnessed a rapid increase in of our understanding about magnetic skyrmions~\cite{Bogdanov_SPJ1989, Zhou_NSR2019, Zhang_JPCM2020, Fert_NRM2017, Kang_PIEEE2016, Nagaosa_NatNano2013, R_Nat2006}. These spin textures are topologically-protected, and can be effectively manipulated by spin currents~\cite{Sampaio_NN2013, Woo_NM2016} or electric field~\cite{Ma_NL2018,Hu_NCM2018}, thus may serve as ideal information carriers. In recent years, the in-plane analogue of a magnetic skyrmion, named a magnetic bimeron, is gaining a lot of attention~\cite{Kim_prb2019,Gobel_prb2019,Yu_NC2018,Gao_NC2019,Kolesnikov_SR2018,Leonov_prb2017,Kharkov_prl2017,Zhang_SR2015,Lin_prb2018,Ezawa_prb2011,Murooka_SR2020,Zarzuela_prb2020}. The topologically non-trivial square meron lattice has been experimentally observed in the chiral magnet Co$_8$Zn$_9$Mn$_3$~\cite{Yu_NC2018}. More recently, isolated meron-antimeron pairs and bimerons have been stabilized in Py film by magnetic imprinting~\cite{Gao_NC2019}. Despite the topological equivalence between the skyrmion and the bimeron soliton, their magnetic static and dynamical properties are distinct, making the magnetic bimeron attractive from the perspectives of both fundamental physics and practical applications in spintronic materials with in-plane anisotropy.

On the other hand, intensive efforts have been devoted to exploit the topological spin textures in novel magnetic systems, such as two-dimensional (2D) materials~\cite{Tong_NanoLett2018}, frustrated materials~\cite{Kharkov_prl2017,Zhang_NC2017}, liquid crystals~\cite{Foster_NP2019,Duzgun_pre2018} and antiferromagnets~\cite{Baltz_RMP2018,Gomonay_NP2018,Jungwirth_NN2016,Barker_prl2016,Dohi_NC2019,Shen_PRL2020,Zhang_SR2016,Shen_PRAppl2019,Legrand_NM2019}. Among them, the antiferromagnetic materials show great potential to bring topological spin textures closer to the real applications. In AFM systems, the magnetic moments of the coupled sub-lattices cancel out, which leads to nearly zero dipolar field and enhances the stability of nanoscale topological structures. The topological charges in the sub-lattice space also cancel out, and thus making the spin textures free from the skyrmion Hall effect~\cite{Zhang_NC2016,Jiang_NP2017,Litzius_NP2017}. Moreover, the canting of the magnetic momentum leads to extra torques facilitating ultrafast dynamics of AFM system. As a result, the mobility of AFM topological structures is much higher than their ferromagnetic (FM) counterparts~\cite{Barker_prl2016,Dohi_NC2019,Shen_PRL2020}. Consequently, the manipulation of AFM topological structures has much lower power consumption, which is highly attractive from the point of view of practical applications.

In this work, we demonstrate the stabilization of asymmetric bimeron solitons and clusters in AFM thin film with interfacial DMI~\cite{Rohart_prb2013}. We combine analytical and numerical approaches, and systematically investigate their statics and current-driven dynamics. Our main result lies in two aspects: (1) AFM bimerons can be effectively manipulated by spin current, and their dynamics, such as velocity and deformation, have a strong dependence on the direction of the current polarization. (2) AFM bimeron solitons exhibit translational attractive interaction, which enables the formation of clusters with a wide range of N\'eel topological numbers. Moreover, their counterparts with opposite N\'eel topological charge exist. The above characteristics envision a rich class of particle-like spin textures allowed by the in-plane AFM system, which can be manipulated by currents with high flexibility and efficiency. On these merits, AFM bimeron clusters may open the avenue for innovative spintronic devices enabling new paradigm of data manipulations.

\vbox{}
\section{Results}
\label{se:Results}
\noindent
\textbf{Theoretical model.} We consider the G-type cubic antiferromagnetic thin film with in-plane uniaxial anisotropy. And the interfacial DMI can be introduced by an adjacent heavy metal layer. The total Hamiltonian of the AFM system can be written as
\begin{equation}
{\cal H} = -\sum_{\langle k,l \rangle}J\boldsymbol{S}_{k}\cdot\boldsymbol{S}_{l}-\sum_k K_{a}(\boldsymbol{S}_{k}\cdot\boldsymbol{n}_{e})^{2}+\sum_{\langle k,l \rangle} \boldsymbol{D}_{kl}\cdot(\boldsymbol{S}_{k}\times\boldsymbol{S}_{l}),\tag{1} \label{eq:1}
\end{equation}
\noindent
where $\boldsymbol{S}_k$ and $\boldsymbol{S}_l$ ($\left|\boldsymbol{S}\right|$ = 1) are the spin vectors, $J$ (< 0 for antiferromagnets), $K_{a}$ and $\boldsymbol{D}_{kl}$ are the AFM exchange constant, the magnetic anisotropy constant and the DMI vector, respectively, and we take $\left| \boldsymbol{D}_{kl} \right| = D_{I}$.  $\boldsymbol{n}_{e}=\boldsymbol{e}_X$ is the direction of the in-plane magnetic easy axis. By linearly combining  $\boldsymbol{S}_{k}$ and $\boldsymbol{S}_{l}$, the net magnetization $\boldsymbol{m} = (\boldsymbol{S}_{l}+\boldsymbol{S}_{k})/2$ and the staggered magnetization (or N{\'e}el vector) $\boldsymbol{n} = (\boldsymbol{S}_{l}-\boldsymbol{S}_{k})/2$ of the sub-lattice pair are defined. Considering the nearest neighboring spins, the AFM energy in the continuum form can be written as:~\cite{Tveten_prb2016}
\begin{equation}
\begin{aligned}
{\cal H}=&\int{dV} \left\{ {\lambda\boldsymbol{m}^{\text{2}}+{A}[(\partial_{x}\boldsymbol{n})^{\text{2}}+(\partial_{y}\boldsymbol{n})^{\text{2}}+\partial_{x}\boldsymbol{n}\cdot\partial_{y}\boldsymbol{n}]}\right.\\
&\left.{+L\boldsymbol{m}\cdot(\partial_{x}\boldsymbol{n}+\partial_{y}\boldsymbol{n})-{K}(\boldsymbol{n}\cdot\boldsymbol{n}_{e})^{2}+w_{D}} \right\},
\end{aligned}
\tag{2}
\label{eq:2}
\end{equation}
where $\lambda = -8J/a\Delta^{2}$, $A = -J/a$ and $L = -4J/a\Delta$ are the homogeneous exchange constant, inhomogeneous exchange constant and parity-breaking constant, respectively. Here $a$ is the lattice constant, and $\Delta = \sqrt{2}a$. $w_{D}=D[n_{z}(\nabla\cdot\boldsymbol{n})-(\boldsymbol{n}\cdot\nabla)n_{z}]$ is the DMI energy density, with the $D$ being the DMI constant~\cite{Rohart_prb2013}. We note that the signs of the anisotropic terms $\partial_{x}\boldsymbol{n}\cdot\partial_{y}\boldsymbol{n}$, $\boldsymbol{m}\cdot\partial_{x}\boldsymbol{n}$ and $\boldsymbol{m}\cdot\partial_{y}\boldsymbol{n}$ are sensitive to the definition of the sub-lattice pair, and will not affect the cubic symmetry of the AFM system. The detailed derivation of Eq.~(\ref{eq:2}) and the symmetry analysis are given in Supplementary Note 1. We use MuMax3 for the later simulations (see Methods), and adopt the following continuum-scale parameters with $A_\textrm{ex} = -A/2 = -1.6475$ pJ/m, $D = 0.36$ mJ/m$^2$, $K = 5.8\times10^{4}$ J/m$^3$, saturation magnetization $M_{s} = 3.76 \times 10^{5}$ A/m, damping constant $\alpha = 0.01$. Taking the lattice constant $a$ = 0.5 nm, the atomic-scale parameters for Eq.~(\ref{eq:1}) can be derived as $J = 2aA_\textrm{ex} = -1.6475 \times 10^{-21}$ J per link, $D_{I} = a^2D = 9\times 10^{-23}$ J per link and $K_{a} = a^3K= 7.25\times 10^{-24}$ J per atom. The above exemplary material system has similar properties to the perovskite structure KMnF$_3$~\cite{Barker_prl2016,Salimath_PRB2020}, with an exchange constant lower than the experimental value~\cite{Pickart_jap1966} to facilitate the formation of the bimeron clusters. On the other hand, we demonstrate that the bimeron soliton and clusters can also exist in the synthetic antiferromagnets composed of Co based alloy (cf. Supplementary Note 6). In addition, similar spin textures have been experimentally observed in the $\alpha$-Fe$_{2}$O$_{3}$ thin film with a Pt over-layer~\cite{Hariom_arxiv2020}. These results suggest the possibility to stabilize bimerons in a wide range of material systems.

\begin{figure}[t]
	\centerline{\includegraphics[width=0.43\textwidth]{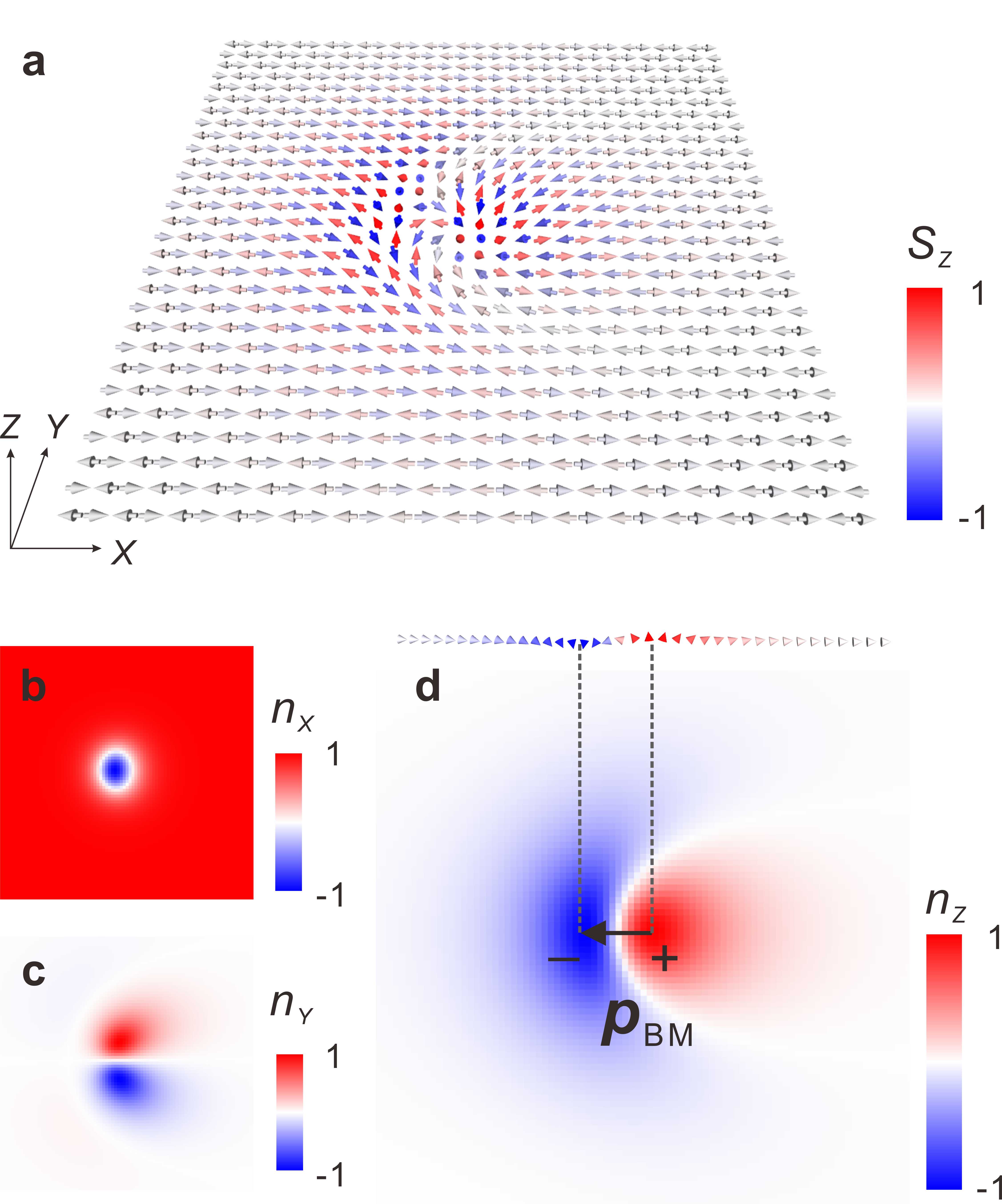}}
	\caption{\textbf{Structure of the AFM bimeron soliton.} \textbf{a} The numerically solved real-space spin texture. The arrows indicate the spin vectors. \textbf{b-d} The N\'eel vector components $n_X$, $n_Y$, $n_Z$.}
	\label{FIG1}
\end{figure}

\vbox{}\noindent
\textbf{Stabilization of AFM bimeron solitons and their current-driven dynamics.} Similar to the AFM skyrmions, the magnetic topology of the AFM bimeron is defined by the N\'eel topological number $Q_\textrm{n} = (1/4\pi)\int{dxdy}[\boldsymbol{n}\cdot(\partial_{x}\boldsymbol{n}\times\partial_{y}\boldsymbol{n})]$, with the uniform ground state $\boldsymbol{n}_X=(1,0,0)$. Figure~\ref{FIG1}a shows the real-space spin texture of the AFM bimeron soliton with $Q_\textrm{n}=+1$, which is formed by a circular AFM meron and a crescent-shaped AFM anti-meron. The corresponding N\'eel vector components are shown in Figs.~\ref{FIG1}b-d, and the structural feature is clearly illustrated by $n_Z$, which indicates a shape-defined magnetic topological dipole. We define the vector pointing from the perpendicular sub-lattices located in the circular meron part to that in the crescent-shaped anti-meron part as the bimeron polarity $\boldsymbol{p}_\textrm{BM}$, which is parallel to the in-plane magnetic easy axis.

Compared with the FM skyrmion, the AFM skyrmion can be manipulated by spin current with higher mobility and no skyrmion Hall effect~\cite{Zhang_NC2016,Jiang_NP2017,Litzius_NP2017}. The AFM bimeron shares similar merits, while exhibiting obvious anisotropic dynamics on the other hand. In this part, we derive the steady motion velocity of the AFM bimeron based on Thiele's approach~\cite{Thiele_prl1973,Tveten_prl2013}, and systematically investigate the dynamics driven by current with in-plane polarization (CIP) and current with out-of-plane polarization (COP). The former case can be realized by utilizing the spin Hall effect in antiferromagnet/heavy metal heterostructures, and the latter case corresponds to the spin-transfer torque originating from the out-of-plane current polarized by a perpendicularly fixed magnetic layer. 

Taking the damping-like spin torques into account, the dynamics of the net magnetization $\boldsymbol{m}$ and the N{\'e}el vector $\boldsymbol{n}$ obey the following two coupled equations~\cite{Hals_prl2011,Velkov_NJP2016}
\begin{linenomath}
	\begin{align}
	\label{eq:3a}
	\boldsymbol{\dot{n}}&=(\gamma\boldsymbol{f}_{m}-\alpha\boldsymbol{\dot{m}})\times\boldsymbol{n}+\gamma H_{d} \boldsymbol{m}\times\boldsymbol{p}\times\boldsymbol{n},\tag{3a} \\
	\boldsymbol{\dot{m}}&=(\gamma\boldsymbol{f}_{n}-\alpha\boldsymbol{\dot{n}})\times\boldsymbol{n}+\boldsymbol{T}_{nl}+\gamma H_{d} \boldsymbol{n}\times\boldsymbol{p}\times\boldsymbol{n},\tag{3b}\label{eq:3b}
	\end{align}
\end{linenomath}
where $\gamma$ and $\alpha$ are the gyromagnetic ratio and the damping constant. $\boldsymbol{f}_{m}=-\delta {\cal H}/\mu_{\text{0}}M_{s}\delta\boldsymbol{m}$ and $\boldsymbol{f}_{n}=-\delta {\cal H}/\mu_{\text{0}}M_{s}\delta\boldsymbol{n}$ are the effective fields.
$\boldsymbol{T}_{nl}=(\gamma\boldsymbol{f}_{m}-\alpha\boldsymbol{\dot{m}})\times\boldsymbol{m}$ is the higher-order nonlinear term~\cite{Hals_prl2011}.
$\boldsymbol{p}$ is the direction of current polarization. $H_{d}=j\hbar P/(2\mu_{0}eM_{s}t_{Z})$ is the equivalent field of spin torque, with $j$ being the charge current density, $\hbar$ the reduced Planck constant, $P$ the spin polarization efficiency, $\mu_{0}$ the vacuum permeability constant, $e$ the elementary charge, and $t_{Z} = a$ the AFM layer thickness.

Based on Eqs.~(\ref{eq:3a}) and~(\ref{eq:3b}), the steady motion speed of the AFM bimeron soliton can be semi-analytically expressed as
\begin{equation}
\begin{pmatrix} v_{x}\\v_{y} \end{pmatrix}=\frac{\gamma H_{d}} {\alpha[d_{xx}d_{yy}-(d_{xy})^{2}]} \begin{pmatrix} -d_{yy} & d_{xy}\\ d_{xy} & -d_{xx} \end{pmatrix}\begin{pmatrix} u_{x}\\u_{y} \end{pmatrix},\tag{4}
\label{eq:4}
\end{equation}
where $d_{ij}=\int{dxdy}(\partial_{i}\boldsymbol{n}\cdot\partial_{j}\boldsymbol{n})$ is the component of the dissipative tensor $\boldsymbol{d}$, and $u_{i} = \int{dxdy}[(\boldsymbol{n}\times\boldsymbol{p})\cdot\partial_{i}\boldsymbol{n}]$ relates to the force induced by the damping-like spin torque~\cite{Thiele_prl1973}. The detailed derivation of the above equation is given in Supplementary Note 2. Equation~(\ref{eq:4}) applies for not only bimeron solitons, but also bimeron clusters with higher $Q_\textrm{n}$, which will be discussed later.


\begin{figure}[t]
	\centerline{\includegraphics[width=0.48\textwidth]{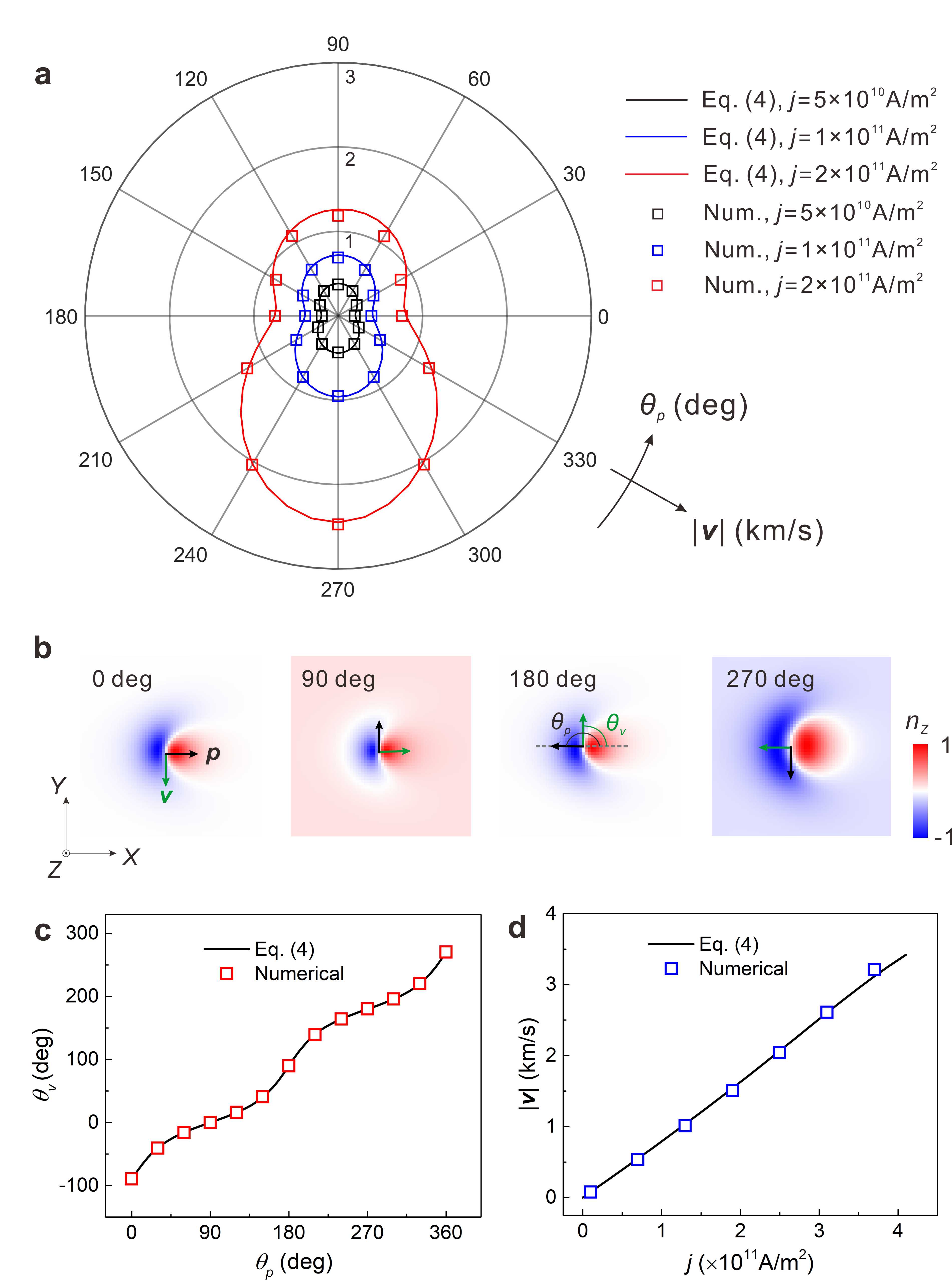}}
	\caption{\textbf{Dynamics of the AFM bimeron soliton excited by spin currents.} \textbf{a} Steady motion speed of an AFM bimeron soliton driven by CIP as a function of the angle $\theta_p$ between the in-plane $\boldsymbol{p}$ and the $+X$ axis. \textbf{b} Deformations of an AFM bimeron soliton induced by CIP with $j = 2 \times 10^{11}$ A/m$^2$, when $\theta_p$ = 0$^\circ$, 90$^\circ$, 180$^\circ$ and 270$^\circ$, respectively. The directions of $\boldsymbol{p}$ and the velocity $\boldsymbol{v}$ are denoted by the black and the green arrows, respectively. \textbf{c} $\theta_\textit{v}$ as a function of $\theta_p$ when CIP with $j = 5 \times 10^{10}$ A/m$^2$ is applied. \textbf{d} Steady motion speed of an AFM bimeron soliton driven by COP as a function of $j$. Data in \textbf{a}, \textbf{c} and \textbf{d} are obtained using both micromagnetic simulations and the semi-analytical approach described by Eq.~(\ref{eq:4}).}
	\label{FIG2}
\end{figure}


One of the key features of the AFM bimeron soliton is its anisotropic dynamics excited by CIP. Figure~\ref{FIG2}a shows the motion speed of the AFM bimeron soliton with $Q_\textrm{n} = +1$ as a function of $\theta_p$, the angle between the in-plane $\boldsymbol{p}$ and the $+X$ axis, with a varying current density $j$. Here the speed of the AFM bimeron soliton is first obtained by numerically tracking its guiding center (cf. Methods), and then comparing with Eq.~(\ref{eq:4}). We assume the spin polarization efficiency $P$ = 0.1, and for a moderate current with $j = 5 \times 10^{10}$ A/m$^2$, the maximum speed is reached at $\theta_p$ = 90$^\circ$ or 270$^\circ$. The one-fold anisotropic dynamics reflect the in-plane nature of the magnetic bimeron, which acquires the largest spin torque when $\boldsymbol{p}$ is orthogonal to the magnetic easy axis. On the other hand, the mobility for $\theta_p$ = 0$^\circ$ or 180$^\circ$ is due to the asymmetric structure of the soliton, which is absent for the symmetric AFM bimeron stabilized by anisotropic DMI~\cite{Shen_PRL2020}. As the current density increases to $1 \times 10^{11}$A/m$^2$, and further to $2 \times 10^{11}$A/m$^2$, the strong spin torques deform the bimeron soliton, leading to the two-fold anisotropic dynamics. In particular, the spin textures of an AFM bimeron for $j = 2 \times 10^{11}$ A/m$^2$ with varying $\theta_p$ are shown in Fig.~\ref{FIG2}b to illustrate the current-induced deformations. A significant constriction/expansion can be observed when $\theta_p = 90^{\circ}/270^{\circ}$, which leads to a lower/higher motion speed of the soliton. On the other hand, only a slight tilt of spin textures is observed when $\theta_p$ = 0$^\circ$ or 180$^\circ$ . Here we note that the deformation of an AFM bimeron soliton involves the change of effective AFM mass tensor  $\boldsymbol{M}_\text{eff}={\mu_{0}}^{2}{M_{s}}^{2}t_{Z}\boldsymbol{d}/\lambda\gamma^2$, as shown in Supplementary Figure 2.  

We next define the motion direction of the AFM bimeron soliton by $\theta_\textit{v}$, the angle between the velocity $\boldsymbol{v}$ and the $+X$ axis. Figure~\ref{FIG2}c shows $\theta_\textit{v}$ obtained by both Eq.~(\ref{eq:4}) and the numerical simulations as a function of $\theta_p$. CIP with a small $j$ is adopted to exclude the influences from the spin torque-induced deformation. The nonlinear dependence of $\theta_\textit{v}$ on $\theta_p$ is reminiscent of the skyrmion Hall effect, which demonstrates a transverse drift of ferromagnetic skyrmions with respect to the charge current flow. For the AFM bimeron soliton, the drifting is due to the asymmetric spin structure rather than the Magnus force, which is absent in the AFM system. It is found that this effect tends to deflect the motion of the bimeron soliton to the direction of the magnetic easy axis.

Another mechanism to drive the AFM bimeron is to utilize the COP, i.e. $\boldsymbol{p}=(0,0,1)$. With this configuration, the spin torque will be exerted on the in-plane topological core of AFM bimeron soliton, and thus lead to a steady motion along the direction $\boldsymbol{e}_Y$. Figure~\ref{FIG2}d shows the speed obtained by Eq.~(\ref{eq:4}) and the numerical simulations with $P = 0.1$. The dynamics excited by COP is relativistic, which is different from the above mentioned CIP case . As the current density further increases, the speed of the AFM bimeron gradually saturates to the maximum spin-wave group velocity, and manifests the Lorentz invariance (cf. Supplementary Figure 3), which has been theoretically demonstrated for AFM skyrmions~\cite{Barker_prl2016,Salimath_PRB2020} and domain walls~\cite{Shiino_PRL2016}. In addition, we note that symmetric magnetic skyrmions are unresponsive to COP~\cite{Sampaio_NN2013}.

\begin{figure*}[t]
	\centerline{\includegraphics[width=0.85\textwidth]{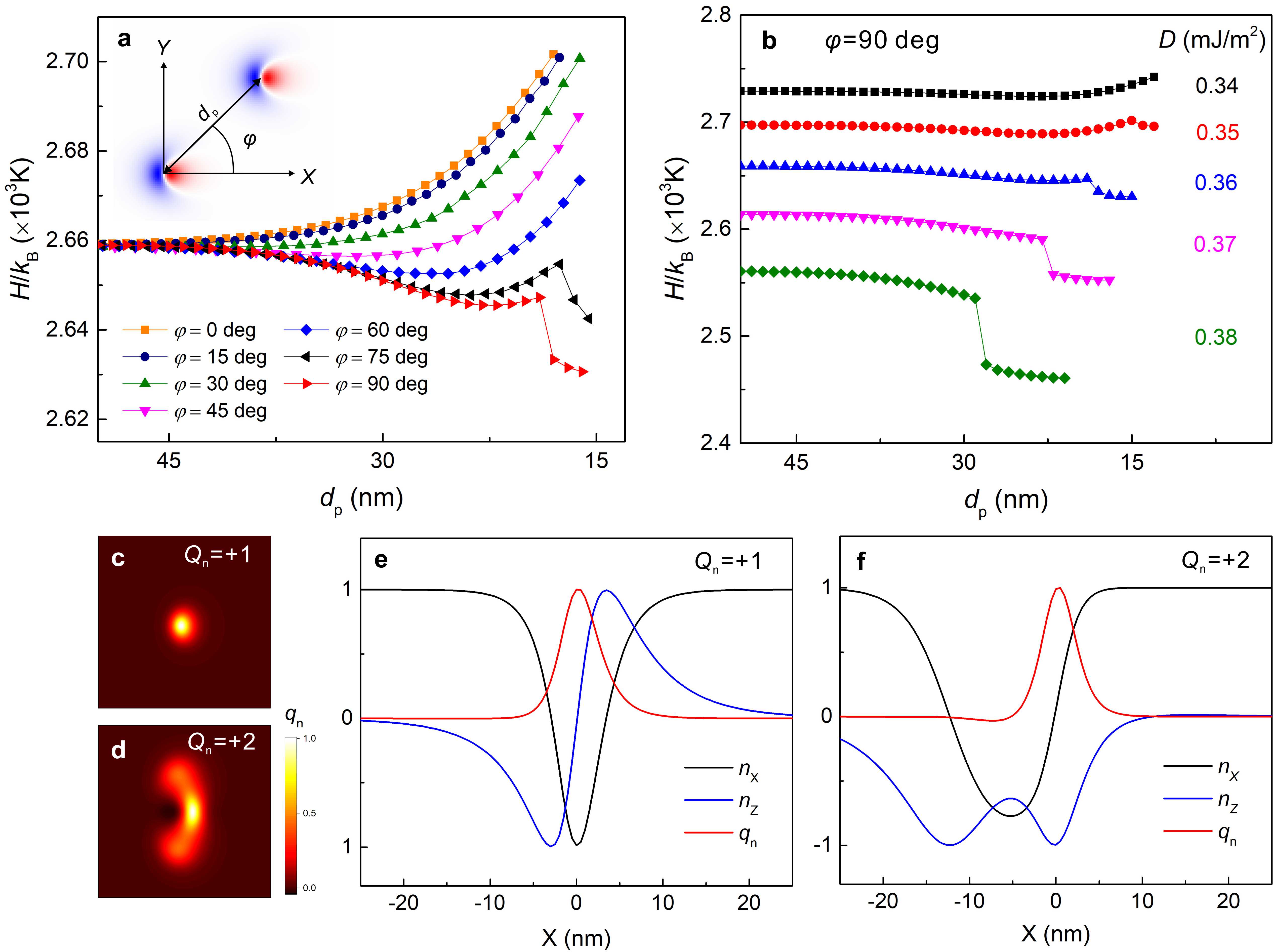}}
	\caption{\textbf{Formation of AFM bimeron dimer.} \textbf{a} System energy as functions of the relative positions of two AFM bimeron solitons. $d_\textrm{p}$ is the distance between the solitons, $\varphi$ is the azimuth with respect to the $+X$ axis, and their definitions are illustrated in the inset. \textbf{b} System energy as a function of $d_\textrm{p}$, with the interfacial DMI constant $D$ varies from 0.34 mJ/m$^2$ to 0.38 mJ/m$^2$. The formation of the AFM bimeron dimer is observed when $D$ > 0.35 mJ/m$^2$, which is indicated by a sudden drop of the system energy. \textbf{c, d} The N\'eel topological density $q_\textrm{n}$ distribution of an AFM bimeron soliton ($Q_\textrm{n}$ = +1) and a dimer ($Q_\textrm{n}$ = +2). \textbf{e, f} The profile of $q_\textrm{n}$, $n_X$ and $n_Z$ along the translational center ($n_Y$ = 0) of the AFM bimeron soliton and dimer. The topological core of the AFM bimeron soliton is featured by in-plane spins with $n_X$ = 1, while that of the AFM bimeron dimer is featured by out-of-plane spins with $n_Z$ = 1.}
	\label{FIG3}
\end{figure*}

\vbox{}\noindent
\textbf{Formation of bimeron clusters.} The bound state of skyrmions has been predicted and observed in chiral ferromagnets~\cite{Zhao_PNAS2016,Yu_NP2018,Gobel_SR2019,Du_prl2018}, frustrated materials~\cite{Kharkov_prl2017,Zhang_NC2017} and liquid crystals~\cite{Foster_NP2019}. Despite the intriguing physics, it is important to find effective methods to manipulate these aggregated topological structures. In this part, we demonstrate the stabilization of the AFM bimeron clusters with high $Q_\textrm{n}$ in chiral antiferromagnetic thin films. The formation of these AFM clusters involves the rearrangement of topological structures, and leads to remarkable changes in both static and dynamical characteristics. Moreover, they have high mobility as generic AFM quasi-particles, making them ideal building blocks for AFM spintronic devices.

We first demonstrate the anisotropic interactions between two bimeron solitons with $Q_\textrm{n}$ = +1, as shown in Fig.~\ref{FIG3}a. For the simulations, we fix the spins at the topological center of each soliton, and minimize the system energy with this constraint. By varying the distance $d_\text{p}$ between the fixed cells along the direction defined by the angle $\varphi$, as illustrated by the inset of Fig.~\ref{FIG3}a, the effective interaction between the solitons can be derived from the variation of the total energy of the system~\cite{Rozsa_prl2016}. For $\varphi$ = 0$^\circ$, the system energy rapidly increases as the bimerons approach each other, demonstrating a repulsive interaction. And the bimerons finally collapse at the separation distance $d_\text{p}$ = 18 nm due to the constriction, as indicated by the orange line. When $\varphi$ increases, the interaction gradually changes from repulsive to attractive. For $\varphi$ = 90$^\circ$, a shallow potential well can be observed at $d_\text{p}$ = 22 nm, as indicated by the red line. Closer approach of the bimeron solitons leads to the formation of a bimeron dimer, which is indicated by a sudden drop of the energy at $d_\text{p}$ = 19 nm. To further investigate the formation of the bimeron dimer, we set $\varphi$ = 90$^\circ$ and vary the strength of DMI, and its influence on the interaction between bimeron solitons is shown in Fig.~\ref{FIG3}b. For $D = 0.34$ mJ/m$^2$, the interaction is repulsive. In this case the energy barrier prohibits the merging of bimeron solitons, and both of them collapse at $d_\text{p}$ = 13 nm. As $D$ increases, the energy barrier vanishes, and finally leads to the spontaneous formation of the bimeron dimer, as indicated by the pink ($D = 0.37$ mJ/m$^2$) and green ($D = 0.38$ mJ/m$^2$) lines. Meanwhile, the merging distance and the bonding energy also increase with $D$. However, for higher $D$, the bimeron soliton tends to extend in the direction perpendicular to the magnetic easy axis, and relaxes to a domain wall pair, as demonstrated in Supplementary Note 4. This phenomenon is similar to the case for skyrmions in the perpendicular AFM system, which narrows the stability region of the solitons~\cite{Bessarab_prb2019}.

\begin{figure*}
	\centerline{\includegraphics[width=0.8\textwidth]{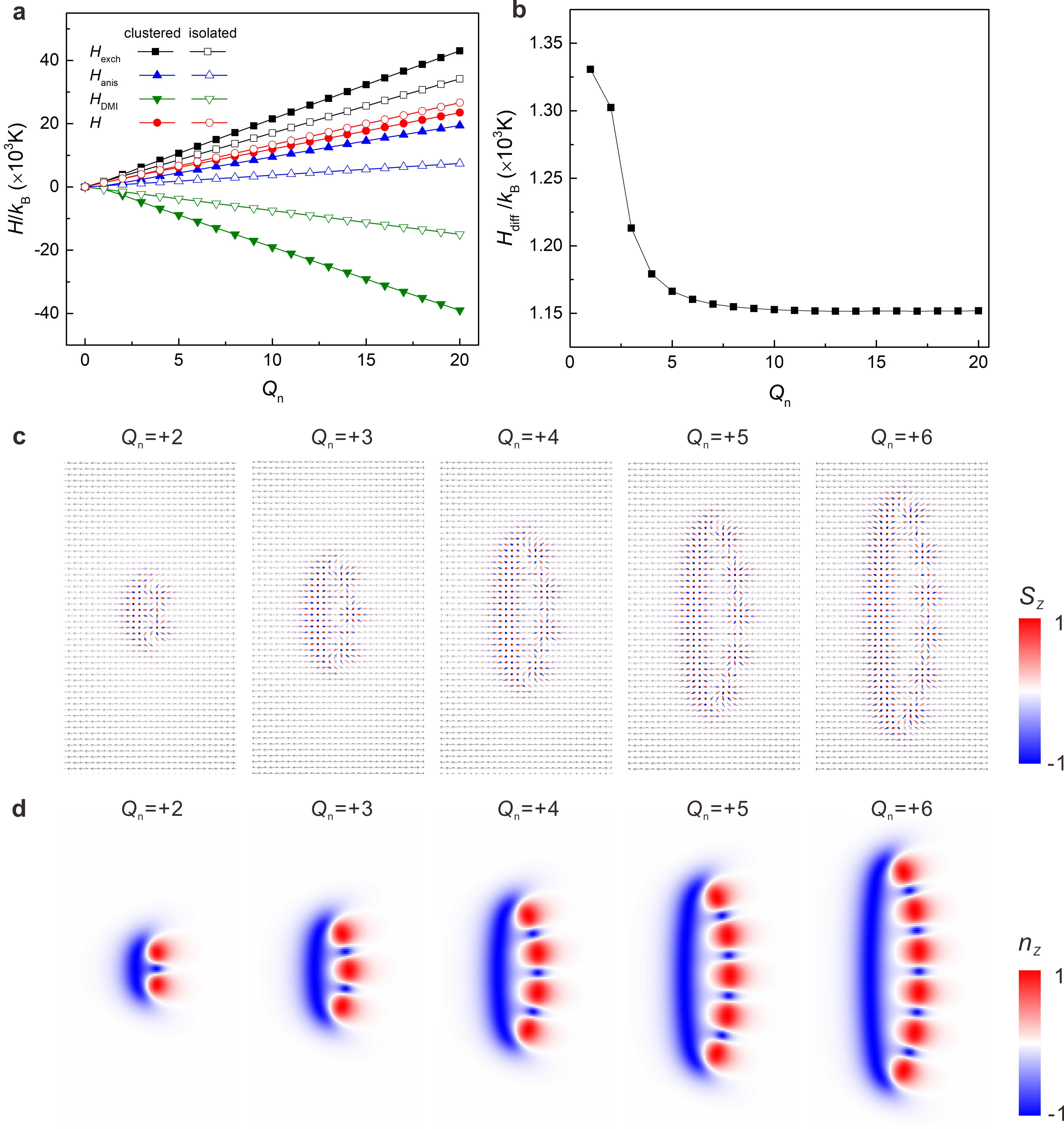}}
	\caption{\textbf{Stabilization of AFM bimeron clusters with high $Q_\textrm{n}$.} \textbf{a} Energy composition of the clustered and isolated bimerons vs. $Q_\textrm{n}$. $\cal{H}_\textrm{exch}$, $\cal{H}_\textrm{anis}$, $\cal{H}_\textrm{DMI}$ are the energy contributions from the AFM exchange, magnetic anisotropy and DMI, respectively. \textbf{b} Energy difference between clusters with neighboring N\'eel topological numbers vs. $Q_\textrm{n}$. \textbf{c} Real-space spin vectors and \textbf{d} $Z$ component of N\'eel vectors of bimeron clusters with $Q_\textrm{n}$ ranging from +2 to +6.}
	\label{FIG4}
\end{figure*}

Figure~\ref{FIG3}c and \ref{FIG3}d show the spatial distribution of normalized N\'eel topological charge density $q_\textrm{n} = \boldsymbol{n}\cdot(\partial_{x}\boldsymbol{n} \times \partial_{y}\boldsymbol{n})$ of the AFM bimeron soliton and dimer, respectively. We note that the formation of a dimer involves the merging of topological cores as shown in Fig.~\ref{FIG3}d, which indicates essential changes of the spin textures. In contrast, the topological cores remain protected for skyrmions in a bound state~\cite{Kharkov_prl2017}. Figures~\ref{FIG3}e and \ref{FIG3}f compare the profiles of N\'eel vector components and $q_\textrm{n}$ along the translational center line ($n_Y = 0$) of the bimeron soliton and dimer. The magnetic topology of bimeron soliton is represented by the in-plane sub-lattice spins ($n_X = -1$), on the other hand, that of bimeron dimer is represented by the out-of-plane ones ($n_Z = -1$). The reorientation of topology-representative spins is another feature identifying the formation of a bimeron dimer.

Due to the translational attractive interaction between the same solitons, they can accumulate along the direction perpendicular to the magnetic easy axis, and thus form AFM bimeron clusters with a wide range of $Q_\textrm{n}$. Figure~\ref{FIG4}a compares the energy composition of bimeron clusters and the isolated bimeron solitons with the same $Q_\textrm{n}$ up to 20. There are two features worth noting. Firstly, DMI prefers the formation of bimeron clusters rather than isolated bimeron solitons, as indicated by the green lines. Secondly, the stabilization of bimeron soliton is mainly determined by the competition between AFM exchange and DMI, while the magnetic anisotropy plays a minor role, as indicated by the lines with hollow symbols. However, the formation of bimeron clusters significantly increases the magnetic anisotropy energy, as indicated by the blue lines. As a result, we note that the stabilization of bimeron solitons and clusters may not be mutually guaranteed. Figure~\ref{FIG4}b shows the energy difference $\cal{H}_\textrm{diff}$ between bimeron clusters with N\'eel topological number $Q_\textrm{n}$ and $Q_\textrm{n}-1$. As $Q_\textrm{n}$ increases, $\cal{H}_\textrm{diff}$ quickly drops and then converges to a value smaller than the energy of an isolated bimeron soliton, which indicates the possibility to stabilize AFM bimeron clusters with an arbitrary $Q_\textrm{n}$. Figure~\ref{FIG4}c and \ref{FIG4}d show the real-space spin textures and $Z$ component of N\'eel vectors of the bimeron clusters with $Q_\textrm{n}$ ranging from +2 to +6. 

\begin{figure}
	\centerline{\includegraphics[width=0.38\textwidth]{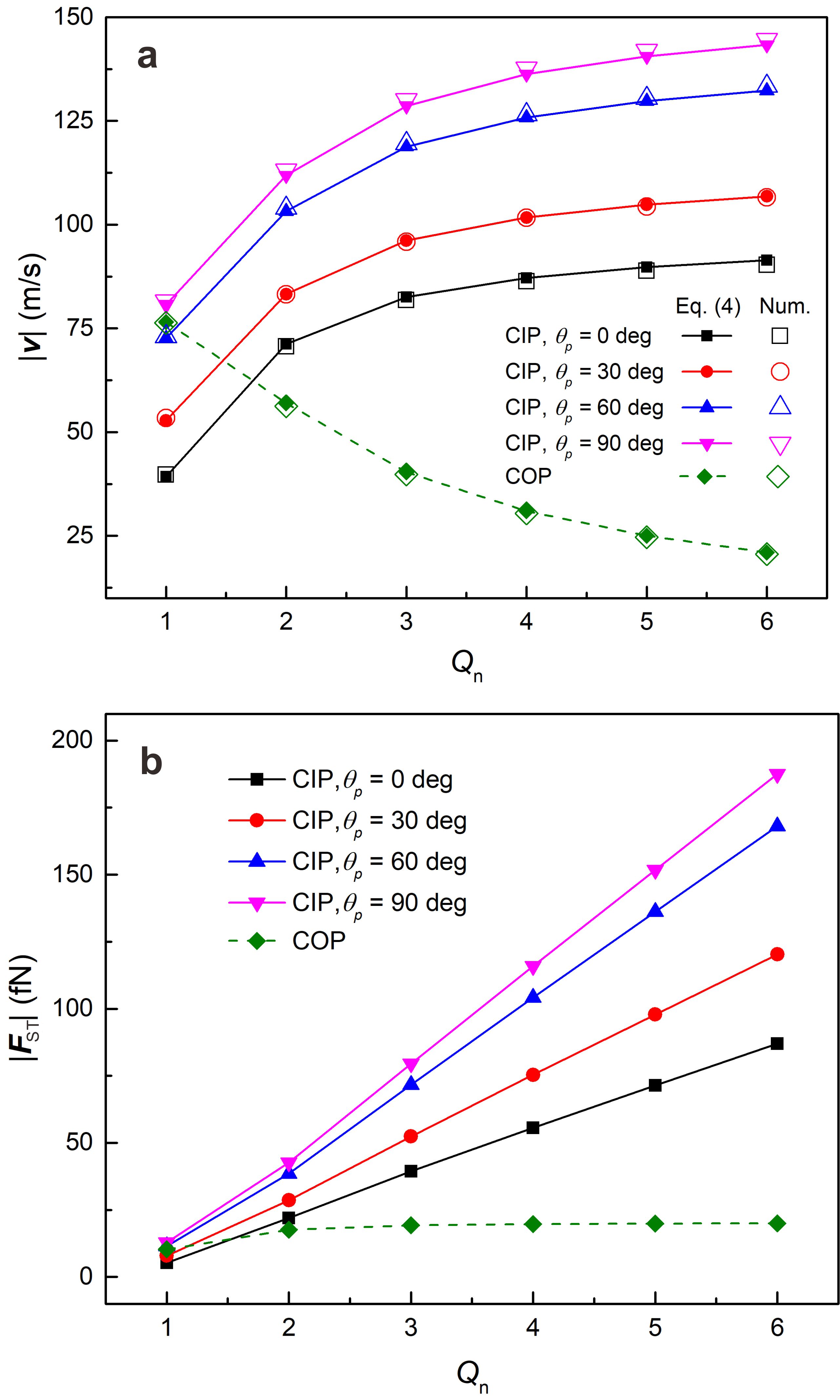}}
	\caption{\textbf{Current-driven dynamics of AFM bimeron clusters.} \textbf{a} The steady motion speeds of the AFM bimeron clusters excited by CIP and COP. The results are obtained using both micromagnetic simulations and the semi-analytical approach described by Eq.~(\ref{eq:4}), with $j = 1 \times 10^{10}$ A/m$^2$ and $P = 0.1$. \textbf{b} The corresponding effective spin torque forces.}
	\label{FIG5}
\end{figure}

\vbox{}\noindent
\textbf{Current driven dynamics of bimeron clusters.} Similar to the solitons, the AFM bimeron clusters can also be effectively manipulated by spin currents. We adopted $j = 1 \times 10^{10}$ A/m$^2$ to avoid the deformation, and the calculated speeds of bimeron clusters driven by CIP and COP are shown in Fig.~\ref{FIG5}a, where good agreements between Eq.~(\ref{eq:4}) and the numerical simulations can be observed. For the CIP-driven case, the AFM bimeron clusters have anisotropic dynamics similar to the soliton, and prefer the motion along the magnetic easy axis. As $Q_\textrm{n}$ increases, the speed driven by CIP nonlinearly increases, as indicated by the solid lines. However, the speed driven by COP decreases, as indicated by the green dashed line. In order to understand this difference, we calculated the strength of the effective spin torque force ($\boldsymbol{F}_{\text{ST},i} = -\mu_{0}H_dM_st_Z\boldsymbol{u}_i$) as a function of $Q_\textrm{n}$, and the results are shown in Fig.~\ref{FIG5}b. For the CIP-driven cases, $\boldsymbol{F}_\text{ST}$ increases almost linearly with $Q_\textrm{n}$, and well explains the anisotropic dynamics observed in Fig.~\ref{FIG5}a. In contrast, for the COP-driven case, the effective force remains constant when $Q_\textrm{n} \geq 2$. Since the accumulation of $Q_\textrm{n}$ will increase the dissipation and the effective mass, but not the effective driving force, the speed of AFM bimeron clusters with higher $Q_\textrm{n}$ tends to decrease.

\vbox{}\noindent
\textbf{Coexistence of topological counterparts and current-driven topology modification.} The topological variety of the bimeron clusters can be further enriched because their counterparts with opposite N\'eel topological number $Q_\textrm{n}$ can coexist within the same AFM background. In order to understand this phenomenon, we analyze the AFM energy and the N\'eel topological number of the bimeron cluster by group symmetry. We use $(S_X, S_Y, S_Z)_A$ to denote the spin vectors of the bimeron A, which are operated as follows,
\begin{equation}
\begin{pmatrix} S_{X}(X,Y)\\S_{Y}(X,Y)\\S_{Z}(X,Y) \end{pmatrix}_{\text{A}} \to \begin{pmatrix} +S_{X}(-X,Y)\\-S_{Y}(-X,Y)\\-S_{Z}(-X,Y) \end{pmatrix}_{\text{B}},\tag{5}
\label{eq:5}
\end{equation}
and then we get the spin vectors of bimeron B. The above operation is performed in the discrete coordinate system (cf. Supplementary Figure 1), and the transformation to the continuum coordinate system leads to
\begin{equation}
\begin{pmatrix} S_{x}\\S_{y}\\S_{z} \end{pmatrix}= \begin{pmatrix} \text{cos}45^{\circ} & \text{sin}45^{\circ} & 0\\ -\text{sin}45^{\circ} & \text{cos}45^{\circ} & 0 \\0 & 0 & 1 \end{pmatrix} \begin{pmatrix} S_{X}\\S_{Y}\\S_{Z} \end{pmatrix},\tag{6}
\label{eq:6}
\end{equation}
The N{\'e}el vector and the net magnetization of the bimeron B are obtained as
\begin{equation}
\begin{pmatrix} n_{x}\\n_{y}\\n_{z} \end{pmatrix}_{\text{B}}= \begin{pmatrix} -n_{y}\\-n_{x}\\-n_{z} \end{pmatrix}_{\text{A}},\begin{pmatrix} m_{x}\\m_{y}\\m_{z} \end{pmatrix}_{\text{B}}= \begin{pmatrix} -m_{y}\\-m_{x}\\-m_{z} \end{pmatrix}_{\text{A}}.\tag{7}
\label{eq:7}
\end{equation}

On the other hand, the operation in Eq.~(\ref{eq:5}) will cause the changes in the spatial derivatives of the AFM N{\'e}el vector,
\begin{equation}
\begin{pmatrix} \partial_{x}\\\partial_{y} \end{pmatrix}_{\text{B}} \to \begin{pmatrix} \partial_{y}\\\partial_{x} \end{pmatrix}_{\text{A}}.\tag{8}
\label{eq:8}
\end{equation}
Based on Eqs.~(\ref{eq:7}) and (\ref{eq:8}), we get
\begin{equation}
\begin{pmatrix} \partial_{x}n_{x}\\ \partial_{x}n_{y}\\ \partial_{x}n_{z} \end{pmatrix}_{\text{B}}= \begin{pmatrix} -\partial_{y}n_{y}\\ -\partial_{y}n_{x}\\ -\partial_{y}n_{z} \end{pmatrix}_{\text{A}}, \begin{pmatrix} \partial_{y}n_{x}\\ \partial_{y}n_{y}\\ \partial_{y}n_{z} \end{pmatrix}_{\text{B}}= \begin{pmatrix} -\partial_{x}n_{y}\\ -\partial_{x}n_{x}\\ -\partial_{x}n_{z} \end{pmatrix}_{\text{A}}.\tag{9}
\label{eq:9}
\end{equation}

Combining Eqs.~(\ref{eq:2}), (\ref{eq:7}) and (\ref{eq:9}), it is found that the AFM energy of the bimeron A is the same as that of bimeron B, while they have the opposite N\'eel topological number $Q_\textrm{n}$. Thus we prove the coexistence of AFM bimeron topological counterparts. Based on a similar approach, we also demonstrate in Supplementary Note 5 that for the clusters with opposite $Q_\textrm{n}$, the CIP leads to their motions in the same direction, while COP to the opposite directions.

\begin{figure}[t]
	\centerline{\includegraphics[width=0.48\textwidth]{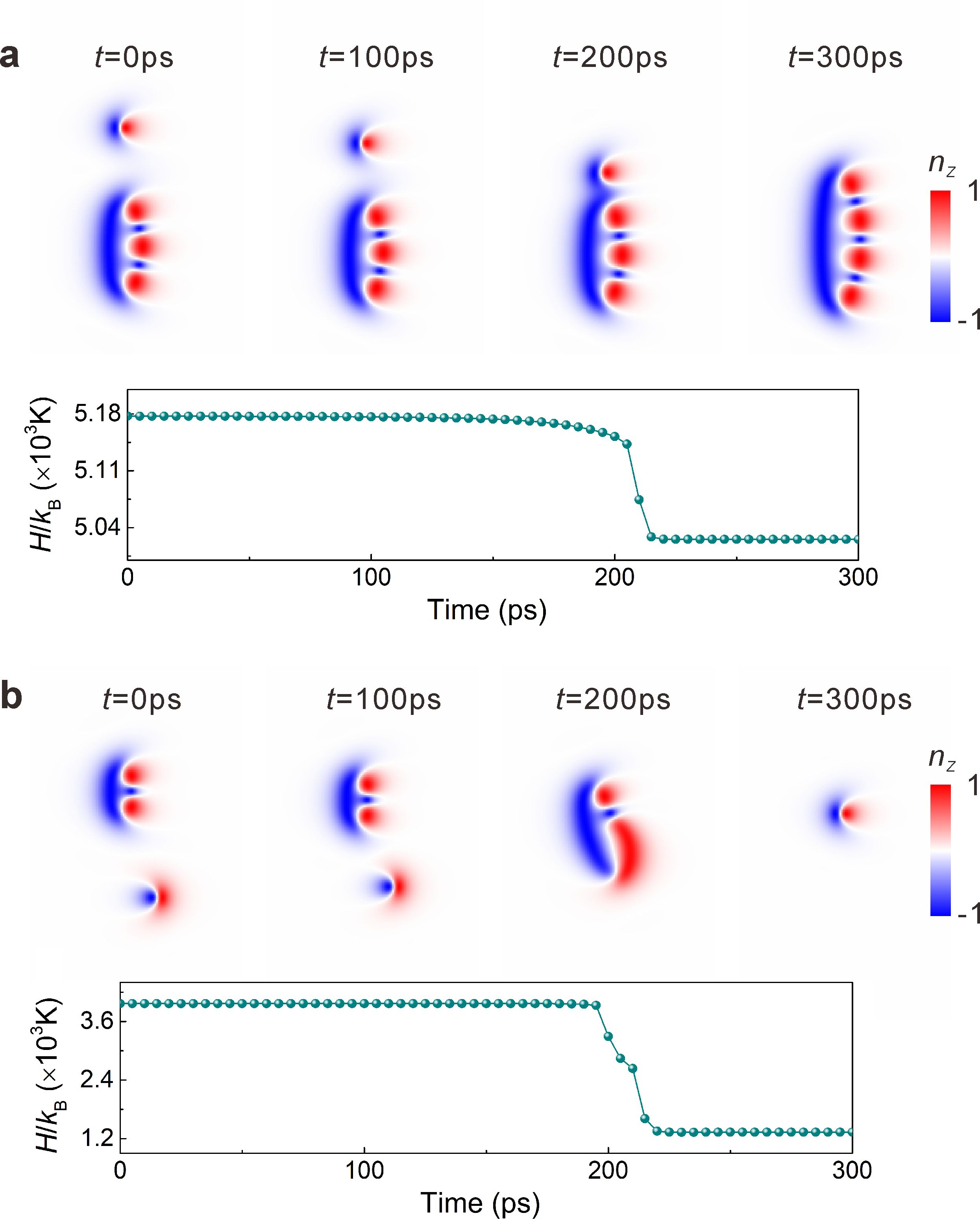}}
	\caption{\textbf{Current-driven topology modification of AFM bimeron clusters.} \textbf{a} Merging of the soliton with $Q_\textrm{n}$ = +1 and the cluster with $Q_\textrm{n}$ = +3. \textbf{b} Annihilation of the soliton with $Q_\textrm{n}$ = -1 and the cluster with $Q_\textrm{n}$ = +2. The upper pannels of \textbf{a} and \textbf{b} are the snapshots of the N\'eel vector distribution at specific simulation time $t$ with an interval of 100 ps, the lower pannels record the system energy during the merging/annihilation process.}
	\label{FIG6}
\end{figure}

Compared with the thin films having perpendicular magnetic anisotropy, the forms of topological quasiparticles allowed by the in-plane AFM system are significantly enriched, making the bimeron clusters ideal multi-bit data carriers. Moreover, through the merging/annihilation of similar/opposite bimeron clusters, their N\'eel topological numbers can be easily modified. Figure~\ref{FIG6} demonstrates the current-driven merging and annihilation process of AFM bimeron clusters. Here we adopted $j = 1 \times 10^{10}$ A/m$^2$, $P = 0.1$, and use COP as the driving force, which leads to the most significant speed differences between clusters with different $Q_\textrm{n}$. To demonstrate the merging of the similar clusters, a bimeron soliton with $Q_\textrm{n}$ = +1 and a cluster with $Q_\textrm{n}$ = +3 are aligned in the direction perpendicular to the in-plane magnetic easy axis. Then COP is applied to drive both the particles in the $-Y$ direction. Due to their speed difference (about 36 m/s according to Fig.~\ref{FIG5}a), the bimeron soliton catches up with the cluster at about 200 ps, and finally they merge to a new cluster with $Q_\textrm{n}$ = +4 (cf. Supplementary Movie 1). After the merging of bimeron soliton and cluster, the total energy of the system decreases by about 160 K. To demonstrate the annihilation of the opposite clusters, a soliton with $Q_\textrm{n}$ = -1 and a cluster with $Q_\textrm{n}$ = +2 are used, and a slight misalignment is introduced to facilitate their annihilation, as shown in Fig.~\ref{FIG6}b. When COP is applied, the soliton and the cluster move toward each other. The annihilation happens at about 200 ps, with a significant drop of system energy (about 2600 K), which leads to a quick burst of spin waves, and finally the soliton with $Q_\textrm{n}$ = +1 remains (cf. Supplementary Movie 2). Based on the varied N\'eel topological numbers, we note that AFM bimeron clusters actually provide a full set of signed integers, and the above-mentioned processes can be regarded as the analogies of summation "3+1" and subtraction "2-1". In this way, the above mentioned process reveals an appealing path towards magnetic topology-based computing.

\vbox{}
\section{Discussion}
\label{se:Discussion}

In summary, this study demonstrates that AFM bimeron clusters with a wide range of N\'eel topological numbers of different sign have the potential to be stabilized in antiferromagnetic thin films with interfacial DMI, and that they exhibit rich and versatile current-driven dynamics. Such findings indicate that AFM bimerons may serve as an ideal candidate to investigate skyrmion-related physics, such as particle interaction, attraction, repulsion, bonding and mutual annihilation. 

From an applied perspective, through the processes of particle merging and annihilation driven by spin currents, the N\'eel topological number of the bimeron clusters can be easily modified, revealing an appealing path towards magnetic topology-based computing. The AFM bimeron cluster may be utilized to unify multi-bits data creation, transmission, storage and computation within the same material system, paving the way for new data manipulation paradigms.

\section{Methods}
\label{se:Method}

\noindent\textbf{Atomistic spin dynamics simulations}
In this work we use the open-source micromagnetic simulator MuMax3\cite{Vansteen_adv} for the modeling of the Heisenberg AFM thin film. While MuMax3 is developed as a micromagnetic simulator for ferromagnetic materials, the finite difference implementation of the Heisenberg-type exchange on the nearest neighbors is also suitable for AFM systems with a simple cubic lattice. By equating the mesh size to the lattice constant, the spin dynamics of the system can be reasonably solved, and the obtained spin structure of the bimeron soliton is identical to that obtained using Vampire\cite{Evans_jpcm2014}, which is an open source atomistic simulator, as shown in Supplementary Figure 4. These results suggest that on the atomistic-scale, the Hamiltonians involved in our system, including the Heisenberg exchange, the magnetic anisotropy and the interfacial DMI, are handled in the same way in MuMax3 and Vampire.

We use currents with in-plane and out-of-plane polarizations to excite the dynamics of the AFM bimerons. In both cases, the currents will induce antidamping N\'eel-order spin-orbit torque on the AFM sub-lattices, with the form $\boldsymbol{S}_{k(l)}\times(\boldsymbol{S}_{k(l)}\times\boldsymbol{p})$\cite{Zelezny_prl2014}. This configuration is self-implemented in MuMax3, and can be directly used without changing the source code. 
Numerically, the guiding center $(r_x, r_y)$ is used to track the position of AFM bimeron soliton and clusters, which is defined by
\begin{equation}
\begin{aligned} 
r_i=\frac{1}{4{\pi}Q_\textrm{n}}\int{i\boldsymbol{n}\cdot(\partial_{x}\boldsymbol{n}\times\partial_{y}\boldsymbol{n})dxdy}, i = x,y.
\end{aligned}
\tag{10}
\label{eq:10}
\end{equation}
And the velocity is calculated by $(v_x, v_y)=(\dot{r}_x,\dot{r}_y)$.

We note that MuMax3 is only used to obtain the spin configuration, based on which the AFM system energy, the topological charge density and other quantities involving magnetization gradient are calculated by self-developed postprocessing tools.

\vbox{}
\section{Code availability}
\label{se:Code availability}
The source code of MuMax3 is available at http://mumax.github.io/. The source code of VAMPIRE is available at https://vampire.york.ac.uk/.



\vbox{}
\noindent\textbf{Acknowledgements}

\noindent
X. L. acknowledges the support by the Guangdong Basic and Applied Basic Research Foundation (Grant No. 2019A1515111110). X.Z. acknowledges the support by the Guangdong Basic and Applied Basic Research Foundation (Grant No. 2019A1515110713), and the Presidential Postdoctoral Fellowship of The Chinese University of Hong Kong, Shenzhen (CUHKSZ). M. E. acknowledges the support from the Grants-in-Aid for Scientific Research from JSPS KAKENHI (Grant Nos. JP18H03676, JP17K05490 and JP15H05854) and the support from CREST, JST (Grant Nos. JPMJCR16F1 and JPMJCR1874). O.A.T. acknowledges the support by the Australian Research Council (Grant No. DP200101027), the Cooperative Research Project Program at the Research Institute of Electrical Communication, Tohoku University, and by UNSW Science International Seed Grant. X. X. acknowledges the support from the National Natural Science Foundation of China (51871137 and 61434002), and the National Key R\&D Program of China (2017YFB0405604). M.M. and M.K. acknowledge support from National Science Center of Poland No. 2018/30/Q/ST3/00416. Y. Z. acknowledges the support by the President's Fund of CUHKSZ, Longgang Key Laboratory of Applied Spintronics, National Natural Science Foundation of China (Grant Nos. 11974298 and 61961136006), Shenzhen Fundamental Research Fund (Grant No. JCYJ20170410171958839), and Shenzhen Peacock Group Plan (Grant No. KQTD20180413181702403). Y. Xu acknowledges the support by the State Key Program for Basic Research of China (Grant No. 2014CB921101, 2016YFA0300803), NSFC (Grants No. 61427812, 11574137), Jiangsu NSF (BK20140054), Jiangsu Shuangchuang Team Program and the UK EPSRC (EP/G010064/1). The atomistic simulations were undertaken on the VIKING cluster, which is a high performance compute facility provided by the University of York. We are grateful for computational support from the University of York High Performance Computing service, VIKING and the Research Computing team.

\vbox{}
\noindent\textbf{Author contributions}

\noindent
Y. Z., X. Z. and L. S. conceived the idea. Y. Z., X. X. and M. K. coordinated and supervised the work. X. L. and J. X. performed the micromagnetic simulation. L. S. and Y. B. carried out the theoretical analysis. J. W., Y. X., R. F. L. E. and R. W. C. performed the atomistic simulation. X. L. and L. S. drafted the study with the input from M. E., O. A. T, M. M. and R. W. C.. All the authors discussed the results and contributed to the manuscript. X. L., L. S. and Y. B. contributed equally to this work.

\vbox{}
\noindent\textbf{Competing interests}

\noindent
The authors declare no competing interests.
\end{document}